\documentstyle [12pt]{article}

\begin{document}

\begin{flushright}
Preprint MRI-PHY/14/94, TCD-7-94 \\
hep-th/9410031 , October 1994.
\end{flushright}

\vspace{4mm}

\begin{center}

{\Large \bf Inconsistent Physics in the  \\
\vspace{2ex}
Presence of Time Machines } \\
\vspace{4ex}
{\large S. Kalyana Rama$^1$ and Siddhartha Sen$^2$ }

\vspace{2ex}

(1) Mehta Research Institute, 10 Kasturba Gandhi Marg,

Allahabad 211 002, India.

(2) School of Mathematics, Trinity College, Dublin 2, Ireland. \\

\vspace{1ex}

email: krama@mri.ernet.in  , sen@maths.tcd.ie \\

\end{center}

\vspace{4mm}

\begin{quote}
ABSTRACT.  We present a scenario in $1 + 1$ and $3 + 1$ dimensional space
time which is paradoxical in the presence of a time machine. We show that
the paradox cannot be resolved and the scenario has
{\em no} consistent classical solution. Since the system is macroscopic,
quantisation is unlikely to resolve the paradox. Moreover, in the absence
of a consistent classical solution to a macroscopic system, it is not
obvious how to carry out the path integral quantisation. Ruling out, by
fiat, the troublesome initial conditions will resolve the paradox, by
not giving rise to it in the first place.
However this implies that time machines have an influence on events,
extending indefinitely into the past,
and also tachyonic communication between physical
events in an era when no time machine existed. If no resolution to the
paradox can be found, the logical conclusion is that time machines of
a certain, probably large, class cannot exist in $3 + 1$ and $1 + 1$
dimensional space time, maintaining the consistency of known physical
laws.

\vspace{1ex}

\end{quote}

\newpage

\vspace{2ex}

\centerline{\bf 1. Introduction}

\vspace{2ex}

The existence of traversible wormholes will lead to travel back
in time if, as described in \cite{t88}, one of the mouths of such a
wormhole has travelled at a high velocity relative to
the other and has come
back to rest. Such a wormhole generates a closed time like curve
(CTC), and hence the possibility of travelling back in time.

The formation and the existence of a traversible wormhole, or a CTC, is
itself not proven beyond doubt. It is not even clear if any viable model
exists in which a CTC appears. Recently, Gott \cite{gott} constructed a
simple model where a CTC appears when two infinitely long, parallel cosmic
strings approach each other at high velocity. However, a complete analysis
of this model \cite{thooft1}-\cite{cutler} reveals that either \\
(i) the region containing the Gott CTC is not local
\cite{thooft1,thooft2,cutler}; or, \\
(ii) there is not enough energy in an open universe to
create a CTC \cite{guth}; or, \\
(iii) in a closed universe, the universe collapses to a zero volume before
any CTC can form \cite{thooft2}; or \\
(iv)the appearance of a CTC leads to a
violation of weak energy condition or a divergent energy momentum
tensor, the back reaction of which will, conceivably, prevent the
formation of a CTC \cite{h}. This situation has led Hawking to propose a
``Chronology Protection Conjecture'', according to which the laws of
physics forbids the apperance of CTCs.

Also, in the creation of a CTC by a traversible wormhole, as in
\cite{t88}, the Cauchi horizon where a CTC forms seems to be stable
classically, but not quantum mechanically \cite{t88,h,kimt}.
Thus, in short, no compelling, physically acceptable
model exists at present which can create locally a ``time machine''.
In the following, we generically use the word time machine to denote
traversible wormhole, or a CTC, or anything that permits a travel back
in time.

The main obstacle in settling these issues is the
lack of understanding of
quantum gravity. Thus, it is not clear how nature forbids, if at all,
the creation of a time machine.
Perhaps, a theory of quantum gravity may answer
this question, but no such theory is in sight in which such calculations
can be undertaken.

However, one can assume that a time machine
exists locally, avoiding completely the question of its formation,
and explore its physical consequences in the classical, semiclassical or
quantum regime. The existence of a time machine often leads to paradoxes
which, if cannot be resolved, imply the non existence of the
time machines themselves, if the known physical laws are required
to be consistent.

This approach has been advocated by ``the Consortium'' consisting of
Friedman et al in \cite{t90}, and is being actively pursued at present
\cite{t91}-\cite{p94}. These authors have studied the
unitarity of evolution, consistency of probabilistic interpretation,
etc.\ in free and interacting quantum field theories, and also in quantum
mechanics \cite{others}-\cite{p94}.  For free quantum field
theories, unitarity and probabilistic interpretation are consistent in the
``chronal region'' - space time region with no CTCs, while for the
interacting field theories the evolution is often non unitary.
Hartle has formulated a generalised quantum mechanics incorporating the
non unitary evolution \cite{hartle}.

An immediate paradox that springs to one's mind, upon hearing the word time
machine, is the science fiction scenario where an observer travels back in
time and kills his younger self. An idealised version of this
scenario is the so called Polchinski's paradox \cite{t90} in which a
billiard ball falls into a wormhole,
emerges at the other end in the past,
and (the situation is so arranged that it) heads off straight for its
younger self, knocking it out of its original trajectory, thus
preventing it from falling into the wormhole in the first place.
This is the main situation
that has been studied extensively in \cite{t90}-\cite{n93}, and is shown
in figure 1. (In these works,  more general situtaions are also considered
that includes inelastic, frictional, non planar collisions, etc.\ .
Novikov has considered different situations which are more or less the
same, in spirit, as the billiard ball collison in Polchinski's paradox.)
The resolution of this paradox is that, the collision between older and
younger ball is not head on, but only a glancing one.
The collision is such that
the younger ball deflects from its original trajectory, but only
slightly, and still falls into the wormhole, and comes out in the past
along a slightly different trajectory, which leads to the glancing
collision instead of a head on one.
This is shown in figure 2. In \cite{t90,t91,n93}, a consistent
solution corresponding to the above description is presented in detail.
Reverting to the science fiction scenario, this resolution can be
described picturesquely, albeit crudely, as follows. The time travelled
observer attacks his younger self, causing only a serious
injury. The younger observer, nevertheless, falls into the
wormhole and travels back in time. Hence the time travelled observer is
seriously injured, which is why he could not kill his younger self!

Of course, in the resolution of Polchinski's paradox as described above,
there is now not one consistent solution, but many; perhaps, infinitely
many. However, this situation still obeys all the physical laws, in the
chronal region,  when quantised using path integral formalism \cite{t94}.
Basically, the billiard ball follows one of the many possible consistent
trajectories, with a probability amplitude given by the quantum
mechanical propagator.

In all the situations analysed so far in \cite{t90}-\cite{n93}, there
is atleast one, often many more, consitent classical solution. The
non uniqueness of the classical solution, however, does not lead to
violation of any physical laws when quantum mechanics is incorporated.
Thus it appears that the existence of time machines does not lead to any
irresolvable paradox and, hence, cannot be ruled by demanding the
consistency of the classical and semiclassical physics.

In this paper, we propose a paradoxical scenario, which to the best of
our ability, is irresolvable, without incorporating some unacceptable
physical laws. We take, as our model for time machine, a traversible
wormhole as in \cite{t90} or the Politzer's time
machine \cite{p92}. In these models, the gravitational effects of the
wormhole are assumed to be felt only inside the wormhole and very near
its mouths outside. The outside space time is otherwise unaffected, which
we take to be Minkowskian, where the usual physical laws are
valid.

We first describe a paradoxical
scenario in $1 + 1$ dimensional space time, where
the irresolvability of the paradox is explicit. This case is
special in many respects and, in particular, has limited implications
for the $3 + 1$ dimensional space time. However, the
scenario can be generalised, with some modifications,
to $3 + 1$ dimensional space time where,
again, the irresolvability of the paradox can be seen explicitly.

Of course, such paradoxical situations can be ruled out by fiat by
forbidding the initial conditions that lead to them.
However, since the initial conditions can be imposed at times arbitrarily
far into the past, ruling them out would imply that any time machine that
could be created in the future will have an influence on events, extending
indefinitely into the past. Moreover, as explained in
the paper, the influence of a time machine on past events also implies
a tachyonic communication between events in the past, when no time machine
existed. We would like to note in this connection that in the
generalised quantum mechanics of Hartle \cite{hartle} also, there exist
the phenomena of the CTCs influencing events in the arbitrary past and
of the tachyonic communication. It was, however, in the regime of quantum
mechanics whereas, in our scenario, such non physical aspects are
necessary even in the classical regime if one rules out by fiat the
troublesome initial conditions leading to paradoxes.

Rather, the simplest way out of this paradox
seems to be to rule out the existence
of time machines, atleast of the types considered here. However, our
scenario is likely to be extendable to a wider class of time
machines, which occupies only a finite spatial region and
modifies the space time manifold only locally, leaving it
unaffected everywhere else. Hence, if the paradox in the scenario proposed
here cannot be resolved, the existence of a large class of time machines
may be ruled out.

The paper is organised as follows. In sections 2 and 3, we describe
our scenario, respectively, in $1 + 1$ and $3 + 1$ dimensional space time
discuss their physical implications. In section 4 is a conlcuding
summary.

\vspace{2ex}

\centerline{\bf 2. $1 + 1$ dimensional case}

\vspace{2ex}

As a model of a time machine in $1 + 1$ dimensional space time, we consider
Politzer's time machine \cite{p92}, shown in figure 3. The space
time coordinates are labelled by $(x, t)$ in the following.
The time machine consists of
the spatial interval between $x = 0$ and $x = \Lambda$ at times
$t = 0$ and $t = T$, suitably identified.
The spatial interval at $t = 0$ is identified with
that at $t = T$ such that
the points $(y, 0_-)$ are smoothly identified with
$(y, T_+)$, and the points $(y, T_-)$ with $(y, 0_+)$, for
$0 \le y \le \Lambda$. The region thus formed
contains CTCs and allows travel back in time. We take the space time to
be flat everywhere else, except very near the time machine.

The propagation of a single partice, say a billiard ball, in Politzer's
time machine is also shown in figure 3. This propagation is consistent
with the physical laws, classical and quantum mechanical
\cite{t94} in the chronal region which is spatially and temporally far
outside the time machine, where there are no CTCs.

One can also consider the propagation of two billiard balls, which is
illustrated in figure 4. The two balls are identical and do not interact
with each other, except for collisions which are assumed to be elastic.
The propagation in the chronal region is consistent classically and
quantum mechanically, as shown in \cite{p92}.

Our scenario is a simple extension of the above case. A crucial
ingredient here is that we consider two {\em different} balls, L and H.
The scenario is set up as follows and shown in figure 5.
The time machine is in the
spatial interval between $x = 0$ and $x = \Lambda$ at
times $t = 0$ and $t = T$, as described before.
Let $m_L$ and $m_H \equiv \mu m_L \; , \; \mu \gg 1$,
be the masses of the balls L and H, and $U_L$ and
$U_H$, their respective initial velocities.
Let L and H be, respectively, at
$(-p', 0)$ and $(p, 0)$ at time $t = 0$.
With no collisions, the ball H would be at
$(- q, T)$ at time $t = T$ and, hence, would not have entered the time
machine. Here $p, p'$,and $q$ are positive and are of
order $\Lambda$, so that the initial trajectories of L and H are
well away from the time machine.
The balls L and H will collide first at
A$(a, t_A)$, where $0 < a < \Lambda$ and $0 < t_A < T$.
After this collision, H will enter the time
machine at $(l, T_-)$ where $0 < l < a$ and, emerge from it at
$(l, 0_+)$ with the velocity $V_H$. From the above data, it follows that
\begin{eqnarray}\label{invel}
U_L & = & \frac{a + p'}{t_1} \; > 0  \nonumber \\
U_H & = & - \frac{p + q + \Lambda}{T} \; < 0  \nonumber \\
V_H & = & - \frac{a - l}{T - t_1} \; < 0  \; \; .
\end{eqnarray}
The seperation between L and H at a large and negative time $t_0$,
when the initial conditions on L and H are assumed to be set,
is given by $d_0 = (U_H - U_L) t_0 + (p + p' + \Lambda)$. Also,
if $\frac{\Lambda}{T} \ll c$, the velocity of light, then all the
velocities  can be taken to be non relativistic, which we assume to be the
case here.

In an one dimensional elastic collision between two particles with masses
$m_1$ and $ \mu m_1$ and initial velocities $U_1$ and $U_2$,
the respective final velocities $V_1$ and $V_2$ are given by
\begin{equation}\label{colln}
\left( \begin{array}{c}
         V_1 \\
         V_2
       \end{array} \right)
= \frac{1}{1 + \mu} \left( \begin{array}{cc}
                             1 - \mu & 2 \mu   \\
                             2       & \mu - 1
                           \end{array} \right)
                    \left( \begin{array}{c}
                             U_1 \\
                             U_2
                           \end{array} \right)        \; \; .
\end{equation}
Thus, the scenario, as described above, can
be easily arranged by choosing
\begin{equation}\label{mhml}
\mu \equiv \frac{m_H}{m_L}
= 1 + 2 \left( \frac{V_H - U_L}{U_H - V_H} \right) \; .
\end{equation}
Note that $\mu > 1$, since $U_H < V_H < 0$ and $U_L > 0$.

Let L and H collide at A acoording to the initial set up.
Then, H will enter the time
machine at $(l, T_-)$ and emerge from it at
$(l, 0_+)$ with the velocity $V_H$. As can be clearly seen from figure 5,
since $V_H < 0$, H is now heading
for another collision with L at B$(b, t_B) \; , t_B < t_A$, which is
to the past of the A-collision. The details after this B-collision
can be easily analysed, but are not necessary since they will not affect
our scenario in any crucial way; in particular, they would not resolve
the paradox described below.

Now it is immediately obvious that the above scenario is paradoxical. It
is inconsistent and cannot take place in any sense that we know of.
If there is A-collision, then there will also be a B-collision. But,
the ball L will be reflected back in the B-collision, making impossible
the A-collision. If A-collision did not occur, the ball H would not have
entered the time machine, and there would not be any B-collision. In its
absence, there must be an A-collision, since that was the initial set up.
The argument now repeats.

This is our paradoxical scenario in $1 + 1$ dimensional space time.
We do not know how the paradox can be resolved. It is assumed to be
possible to set the initial conditions on a physical system far in the
past of time machine. However, the evolution of the system
is paradoxical in the presence of a time machine.
It is not clear how both the collisions can take place
nor, how any one or both of them can be avoided. If, for a moment, we
assume that somehow both the collisions {\em did} take place, then it is
easily seen that in the far future of the time machine, there will be one
H ball and {\em two} L balls. This will violate the principle of
conservation of energy, since a new L ball has appeared. Such creation
of real, onshell particles also occurs in interacting quantum field
theories in non causal regions, as shown by Boulaware in
\cite{qftothers}. There, such creation leads to non unitarity of the
field theory.

Now, one can declare that in the presence of a time machine, certain
initial conditions for a physical system, such as the one necessary above,
are forbidden. The known physical laws cannot
accomodate and enforce this kind of censorship,
but let us anyway proceed with
this declaration. This implies, first of all, that any time machine that
could be created in the future will have an influence on the events,
extending indefinitely into the past.

The influence of the time machine on arbitrarily past events also implies
a tachyonic communication between them, at a time when no time machine
existed. Consider the scenario described above, where the balls L and H
were given velocities $U_L$ and $U_H$ initially at a time $t_0$. The
seperation between L and H is
$d_0 = (U_H - U_L) t_0 + p + p' + \Lambda$,
which can be arbitrarily large. If the
time machine to be created at $t \simeq 0$ were to forbid the initial
conditions, such as above, then L and H cannot be given the above
velocities at time $t_0$, which can be any arbitrary time in the past.
Thus, if a team, say L-team,
had set the ball L moving at a velocity $U_L$,
it will be known instantaneously to a H-team
at a distance $d_0$, because this team will not be able
to set the ball H at a velocity $U_H$. This communication must
take place nearly
instantaneously over an arbitrarily large distance $d_0$ and, hence,
can only be tachyonic.

It is not likely that the quantisation of the above system will resolve
the paradox. For one thing, the system involved is macroscopic and its
classical action $S \gg \frac{h}{2 \pi}$, the quantum action.
Hence, although quantisation can resolve the problem of non unique
classical solutions, as in \cite{t94} for example, it is unlikely to
resolve the above paradox in a macroscopic system,
where there is {\em no}  classical solution at all to begin with.
Thus, if a quantum mechanical solution does exist for
the macroscopic scenario described here,
it must be quite unusual, having no classical counterpart.

In the presence of time machines, it has been recognised that Hamiltonian
formulation of quantum mechanics is not possible, since it requires the
space time manifold to be foliable along
the time direction. However, the
path integral formulation is a natural choice in such situations, which
has been successfully employed in the presence of time machines, where
there was atleast one classical solution for macroscopic systems.
But, in our scenario {\em no} consistent classical solution exists, and
following the system along {\em classical}  paths leads to
inconsistencies. It is not obvious how path integral formulation can
be carried out, if possible at all.
Of course, it is possible if, by fiat, all the troublesome
initial conditions are forbidden, but this implies that
time machines have an influence on events extending indefinitely into the
past, and also implies tachyonic
communications between physical events in an era when no time machine
existed. If the above paradox cannot be resolved, then
the logical conclusion is that time machines cannot exist,
while maintaining the consistency of known physical laws.

A major shortcoming of the above scenario is that it is in $1 + 1$
dimensional space time, which has limited implications, if any,
to the $3 + 1$ dimensional space time. However, as we will show in the
next section, this scenario, with a few modifications,
can also be extended to $3 + 1$ dimensional space time, leading to the
same conclusions as above.

\vspace{2ex}

\centerline{\bf 3. $3 + 1$ dimensional case}

\vspace{2ex}

As a model of a time machine in $3 + 1$ dimensional space time, we consider
the time machine created by a traversible wormhole
which is shown in figure 6. Note that, in this
figure, only the X-Y plane of the three dimensional space is shown, in
contrast to previous figures where both the time and the one dimensional
space were shown. The two wormhole mouths, W1 and W2,
of size $\omega$, are centered  respectively at $(0, h)$ and
$(\Lambda, h)$ in the X-Y plane. Thus they are seperated from each other
by a distance $\Lambda$.
One travels back in time, by an amount $T$, by entering
the mouth W2 and emerging from the other mouth W1.
See \cite{t88,t90,t91} for details. The gravitational effects of the
wormhole are assumed to be
felt only inside its traversible throat and very near its
mouths outside. The space time outside is otherwise unaffected, which we
take to be Minkowskian. In the following we assume that
$R \ll \omega \ll \Lambda$, where $R$ is the size
of the physical object traversing the wormhole, and that the
back reaction of the physical system on the wormhole is negligible.

We consider a light ball L of radius $r$ and mass $m_L$,
and a heavy disc D of radius $R \gg r$ and mass
$m_D \equiv \mu m_L, \; \mu \gg 1$.
Thus, if L collides with D away from the edge of D,
then L will be reflected back.
L and D can roughly be thought of as a ping pong ball and a bat.

Initially, at a time $t_0$ which is large and negative, L and D are given
the velocities $(U, 0)$ and $(- V_x, V)$,
respectively, as shown in figure 6. The centers of L and D are in the
X-Y plane. The disc D has only a translational motion
with its normal parallel to the X-axis. As before, the initial conditions
are such that, in the absence of L, the disc D would not enter the
wormhole. With L present, however, there will be a
collision between L and D, with zero impact parameter, at the point
$A = (\Lambda, 0)$ at time $t = t_A = \frac{\Lambda}{U}$.
We choose
\[
V_x = \frac{2 U}{\mu -1}
\]
such that after the collision, D is deflected from its path,
travels with a velocity $(0, V)$, and
falls into the wormhole mouth W2. D will then emerge from W1 with
a velocity $(0, - V)$ travelling along the Y-axis,
and collide with L with zero impact parameter at the point
$B = (0, 0)$ at time $t = 0 < t_A$, {\em i.e.}\ to the past
of the A-collision.
The wormhole traversal rules that we used here are given in \cite{t91}.
According to their notation, if
$\vec{V}_{in} = V (\cos (\theta + \phi), \sin (\theta + \phi))$, then
$\vec{V}_{out} = V (\cos (\theta - \phi), \sin (\theta - \phi))$.
See figure 4 of \cite{t91} for further details. In the set up here,
$\theta = \phi - \frac{\pi}{2} = 0, \;
\vec{V}_{in} = (0, V)$ and,
therefore, $\vec{V}_{out} = (0, - V)$.

There is a paradox here. If the A-collision took place as planned,
then the B-collision will also take place.
But this will reflect back the ball
L at time $t = 0 $, making the A-collision at time $t_A > 0$ impossible.
This scenario is the $3 + 1$ dimensional analog of that of section 2.

But, there can be consistent solutions to this scenario
with no paradox, if as in \cite{t90,t91,n93}, one also considers
not only head on collisions with zero impact parameter, but also the edge
on collisions with non zero impact parameter. However, we will
make a simple modification in the above scenario, in which the disc D,
after emerging from W1, is slowed down  before it encounters L
from a velocity $(0, - V')$ to $(0, - \tilde{V})$, where
$\tilde{V}$ is very small.

Slowing down D is not a problem. It can be done, for example, by having a
viscous medium in between, and well away from, W1 and the collision
point B; or, by having a charged disc D and then applying an electric
field. This process can also be used to make the disc travel
along the Y-axis with its normal parallel to
the X-axis - ``funnelling or steering'' D along the required
trajectory - at the time of its encounter with L. Note that $\tilde{V}$
can be arbitrarily small and that the slowing down process is totally
independent of the wormhole, neither affecting the other. We parametrise
the slow down by a parameter $\alpha \; , \; 0 < \alpha < 1$, saying that
D, after emerging from W1, travels a vertical distance $\alpha h$ with the
speed $V'$ and the remaining distance $\beta h \equiv (1 - \alpha) h$
with the speed $\tilde{V}$. This parametrisation is convenient and quite
general, and does not imply that the entire slowing down process has
occured at one particular location between W1 and B.

Before modification, the scenario is very likely to be consistent
upon including also the edge on collisions between L and D.
They will not be of much help in the scenario, as will be clear below.
The initial set up, as before, is
such that if the collision at $A = (0, \Lambda)$
took place as planned, then there will be a head on collision
at $B = (0, 0)$ leading to an inconsistency. Note that
B-collision always takes place at $t = 0$, since that is when the ball L
will arrive at B. The planned collision at A will occur at
$t = t_A = \frac{\Lambda}{U} > 0$. Any potentially consistent solution
will involve an edge on collision at B at time $t = 0$, and the
A-collision
will now take place at a location $A' = (A_x, d - \frac{\Lambda V}{U})$
at a time $t_{A'} \simeq t_A \pm \frac{R}{\tilde{V}}$.
It must take place somewhere in a finite region
after D is near enough to W2 to make it to B in time, but
before D passes W2. The idea is to choose $\tilde{V}$
sufficiently small and, hence $\frac{R}{\tilde{V}}$ sufficiently large, so
that the disc D will be outside this finite region.

The required condition for the collisions at A and B to occur as planned
is that
\begin{equation}\label{headon}
\frac{\Lambda}{U} + \frac{h}{V} - T + \frac{\alpha h}{V}
+ \frac{\beta h}{\tilde{V}} = 0   \; ,
\end{equation}
where, starting from B, the first term is the time
required for L to reach A, the second term for D to reach W2 from A,
$T$ the amount of time travel,
the last two terms are the time required for D to reach B from W1. The
travel time through wormhole throat is neglected; if non zero, it can be
incorporated in the definition of $T$. Equation (\ref{headon}) implies
\begin{equation}\label{v}
V = \frac{(1 + \alpha) h}{T - \frac{\beta h}{\tilde{V}}
- \frac{\Lambda}{U}} \; .
\end{equation}
Since $V > 0$, it follows that
\begin{equation}\label{u}
U > \frac{L}{T - \frac{\beta h}{\tilde{V}}} \; .
\end{equation}
Furthermore, since $U > 0$, it follows that
\begin{equation}\label{vtilde}
\tilde{V} > \frac{\beta h}{T}  \; .
\end{equation}
The initial velocity of D is $(- \frac{2 U}{\mu - 1}, V)$. Therefore,
\begin{equation}\label{l}
l = \frac{2 U (T - \frac{\beta h}{\tilde{V}}
- \frac{\Lambda}{U})}{(\mu - 1) (1 + \alpha)}  \; ,
\end{equation}
where we have used equation (\ref{v}), and $l$ is the distance
from W2, parallel to the X-axis, at which D would have passed
if there were no collisions.
$l$ is required to be $> \omega$, the wormhole size.

Now for the edge on collision. As noted before, the B-collision still
occurs at $(0, 0)$ at time $t = 0$, but now the A-collision occurs at
$A' = (A_x, d - \frac{\Lambda V}{U})$, instead of at
$A = (\Lambda, 0)$ as planned. Note that the Y-coordinate of
(the younger) D at time $t = 0$ is $- \frac{\Lambda V}{U}$ and, hence,
it would travel a vertical distance of $d > 0$ with speed $V$,
before colliding with L at $A'$. Let the velocity of D after
$A'$-collision
is $(V_x', V')$, and in a consistent solution it will enter W2 and
emerge from W1 with a velocity $(V_x'', V'')$, which is close to
$(0, - V')$. Let $2 \chi$, where $\chi$ is small, be the angle between
the trajectory of D emerging from W1 and the negative Y-axis
(see \cite{t91} for ``wormhole traversal rules'').

The time required for D to enter W2 after $A'$-collision is
\[
(h - d + \frac{\Lambda V}{U}) \frac{1}{V'} \; .
\]
It will emerge from W1 with
a velocity $(V_x'', V'')$, be slowed down and steered along a vertical
trajectory with a velocity $(0, - \tilde{V})$. We write the travel time
from W1 to the collision point at B as
\[
\frac{\alpha h}{V'}  + \frac{\beta h + \sigma R}{\tilde{V}} + \tau \; ,
\]
taking  $(V_x'', V_y'') = (0, - V')$, and
parametrising the necessary corrections by $\tau$, which is of the order
\begin{equation}\label{tau}
\tau \simeq \frac{2 h \tan^2 \chi}{V' (1 - \tan^2 \chi)} \; .
\end{equation}
Also, $\sigma = + 1 \; ( - 1 )$ denote the upper (lower) edge
collision between L and D.

Thus the condition for the edge on collision is,
in the same way as before,
\begin{equation}\label{edgeon}
\frac{d}{V} + (h - d + \frac{\Lambda V}{U}) \frac{1}{V'}
- T + \frac{\alpha h}{V'}
+ \frac{\beta h + \sigma R}{\tilde{V}} + \tau = 0   \; .
\end{equation}
Subtracting (\ref{edgeon}) from (\ref{headon}), we have
\begin{equation}\label{bound}
\frac{\sigma R}{\tilde{V}} =
\left( \frac{1}{V} - \frac{1}{V'} \right)
\left( \frac{\Lambda V}{U} + (1 + \alpha) h - d \right) - \tau    \; ,
\end{equation}
which must be satisfied for any consistent solution.
$V'$ is the Y-component of D after a collision at $A'$ with L. Since D
is much heavier than L, its final velocity cannot be too different from
its intial velocity. Hence,
$\left( \frac{1}{V} - \frac{1}{V'} \right)$ is small and, in particular,
is bounded. Therefore,
the right hand side of equation (\ref{bound}) is also bounded. Thus, if
$\tilde{V}$ can be choosen sufficiently small so as to exceed this bound,
then no consistent solution can exist.

Let $\frac{\beta h}{\tilde{V}} = (1 - 2 \epsilon) T$ and
$\frac{\Lambda}{U} = \epsilon T$, with $\epsilon > 0$. Then,
\begin{equation}\label{vl}
V = \frac{(1 + \alpha) h}{\epsilon T} \; , \; \;
l = \frac{2 \Lambda}{(1 + \alpha) (\mu - 1)} \; .
\end{equation}
Equation (\ref{bound}) now becomes
\begin{equation}\label{kbound}
\frac{\sigma (1 - 2 \epsilon) R}{\beta h} = \epsilon K \; ,
\end{equation}
where
\begin{equation}\label{k}
K \equiv \left( 1 - \frac{V}{V'} \right)
\left( 2 - \frac{d}{(1 + \alpha) h} \right)
- \frac{2 V \tan^2 \chi}{V' (1 + \alpha) (1 - \tan^2 \chi)} \; ,
\end{equation}
where we used (\ref{tau}) as an estimate of $\tau$.

Consider $\frac{V}{V'}$. $V'$ is the Y-component of the
velocity of D after $A'$-collision with L, and is largest when all the
available energy in the collision is converted into D's velocity
in the Y-direction. Of course, this can never happen in any collision
between L and D, but will give an upper bound, although a grossly
overestimated one. Before the $A'$-collision, the velocity of the D was
$(\frac{2 U}{\mu - 1}, V)$. The ball L had an edge on collision at B in
which L was incident horizontally with a velocity $U$, and D with a small
vertical velocity $\tilde{V}$. Since the B-collision was edge on, it only
deflected the original trajectory of L with, to a very good approximation,
a negligible change in the speed of L.
Hence, the energy of L prior to $A'$-collision is
$\frac{m_L U^2}{2}$. If after this collision, all the available energy is
converted into $V'$, then
\[
\frac{1}{2} m_D V'^2 = \frac{1}{2} m_D (V^2 + \frac{4 U^2}{(\mu - 1)^2})
+ \frac{1}{2} m_L U^2 \; .
\]
 From the above, we get
\begin{equation}\label{v1}
1 - \frac{V}{V'} = \frac{U^2}{2 \mu V^2} \;
\left( \frac{\mu + 1}{\mu - 1} \right)^2 + \cdots
\end{equation}
where $\cdots$ denote higher order terms in
$\frac{U^2}{2 \mu V^2} \; \left( \frac{\mu + 1}{\mu - 1} \right)^2$,
which are negligible.

Now consider the $A'$-collision in the other extreme when the disc D
looses maximum possible energy, decreasing its Y-component velocity to a
maximum extent which, again, is a gross overestimate.
In a collision between a light and a heavy body,
the maximum energy is exchanged when the light body is initially at rest
and the collision is head on. The light body gains an amount of energy
$= \frac{1}{2} m_L (2 U_H)^2$, where $m_L$ is its mass and $U_H$, the
initial velocity of the heavy body. Thus in the $A'$-collision in our
case, the maximally reduced $V'$ is given by
\[
\frac{1}{2} m_D V'^2 = \frac{1}{2} m_D V^2 - \frac{2}{\mu}
m_L (V^2 + \frac{4 U^2}{(\mu - 1)^2}) \; ,
\]
from which we get
\begin{equation}\label{v2}
1 - \frac{V}{V'} = - \frac{2}{\mu}
(1 + \frac{4 U^2}{(\mu - 1)^2 V^2}) + \cdots \; ,
\end{equation}
where $\cdots$ denote higher order terms in
$\frac{2}{\mu} (1 + \frac{4 U^2}{(\mu - 1)^2 V^2}) $, which are
negligible. From equations (\ref{v1}) and (\ref{v2}) we see that
$\frac{V}{V'} \simeq 1 + {\cal O}(\frac{1}{\mu})$ atleast and, therefore,
$K$ in equation (\ref{k}) is bounded in magnitude:
\[
| K | \le K_{max} \simeq {\cal O}(\frac{1}{\mu}) \; .
\]

Hence, equation (\ref{kbound}) can now be written as
\begin{equation}\label{kmaxbound}
\frac{(1 - 2 \epsilon) R}{\beta h} \le \epsilon K_{max} \; ,
\end{equation}
which must be satisfied for any consistent solution. However, $\beta$ is
an arbitrary variable in equation (\ref{kmaxbound}) which we are free
to choose. Let
\begin{equation}\label{beta}
\beta = \frac{(1 - 2 \epsilon) R}{2 \epsilon h K_{max}} \; .
\end{equation}
With this choice of $\beta$, equation
(\ref{kmaxbound}) cannot be satisfied and, hence, there can be
no consistent solution. This would imply that the paradox of the
scenario presented here is irresolvable, since no consistent solution
exists. Such a choice of $\beta$ is very much possible in
principle and, in fact, is quite reasonable as we will now
illustrate with specific numerical values for the various
quantities encountered in the present scenario.

Let
\[
\mu = 101 \; , \; \;
\frac{U}{V} = 10 \; , \; \;
K_{max} = 0.01
\]
and
\[
U = 10 m/sec \; , \; \;
R = 1 m \; , \; \;
\omega = 10 m \; , \; \;
\Lambda = 10^4 m \; .
\]
It will turn out that $\alpha \simeq 1$ and $\epsilon \simeq 0$,
so we take $1 + \alpha \simeq 2$ and
$1 - 2 \epsilon \simeq 1$ in the following
estimates. From the above, and using equation (\ref{beta}), we get
\[
V_x = 0.2 m/sec \; , \; \;
V = 100 m/sec \; , \; \;
h = 5 \times 10^4 m \; , \; \;
l = 100 m \; ,
\]
and also
\[
\epsilon T = 1000 \; , \; \;
\epsilon \beta \simeq 0.001 \; ,
\]
from which follows
\[
\tilde{V} = 0.05 m/sec \; .
\]
If we further choose $\epsilon = 0.01$, we have
\[
\beta = 0.1 \; , \; \;
T = 10^5 sec \simeq 30 hrs  \; .
\]

Thus, we have a wormhole with mouths of size $10 m$, seperated by
$10 Km$, allowing travel back in time by
$10^5 sec \simeq 30 hrs$. We also have
a ball of small radius, say of radius $0.1 m$, travelling initially with
a velocity of $(10, 0) m/sec = (36, 0) km/hr$
a disc of radius $1 m$ and $100$ times
heavier than the ball, travelling initially with a velocity of
$(0.2, 100) m/sec = (0.72, 360) km/hr$ in the X-Y plane.
Its initial trajectory would miss
the wormhole by a distance of $100 m$. However, the disc collides with the
ball, and enters the wormhole
and emerges from it $\simeq 30 hrs$ behind in time. After which, the disc
gets slowed down to a crawling velocity of
$(0, 0.05) m/sec = (0, 0.18) km/hr$, and collides
again with the ball in the past, making the first collision impossible,
as shown above.

The above numbers admittedly span a wide range of velocities,
and the set up perhaps requires a high ballistic precision.
However, in principle, nothing forbids the above set up and the numbers.
Also, these numbers have such a wide range because of our deire
to keep the analysis as simple as possible, yet be able to demonstrate
clearly the irresolvability of the paradox in the scenario proposed here.
These numbers can be made more amenable by an elaborate analysis
of the scenario, perhaps aided by numerical simulations,
still resulting in an irresolvable paradox with no consistent solution.

The crucial ingredient in the scenario here is not as much the slowing
down of D to a small velocity $(0, \tilde{V})$, which only made the
analysis easier to handle. Rather, it is that the disc is heavier and
larger in size than the ball so that, in a collision, the ball L will
be reflected back by the disc, unless the collision is edge on. If this
were not the case, then $R$ in the relevent equations above must be
replaced by a variable $\rho$ with a range $0 \le \rho \le R$.
In particular, $\rho$ can be sufficiently small so as to ensure
that equations (\ref{kbound}) and (\ref{kmaxbound}) are always satisfied,
no matter how small $\tilde{V}$ or, equivalently, $\beta$ is. This would
be the case if, for example, both
the colliding objects were identical,
as in the works of \cite{t90,t91,n93}. We would also like to note
that if one can find a situation, which we are unable to, where the
collision between two objects forbids, atleast classically, the
forward scattering of one of them
in a certain finite range, then this situation can also be used to
construct a paradoxical scenario, along the lines of the
present one. For example, in a collision between L
and D in the above scenario, if the forward scattering of L
is forbidden in the range
\[
\arctan \frac{h}{\Lambda} < \theta_s < \arctan \frac{V}{U} \; ,
\]
where $\theta_s$ is the forward scattering angle of L, then a little
thought makes it immediately clear that the paradox of
the scenario described above just cannot be resolved.

The implications of the irresolvability of the paradox in the scenario
presented here is the same as those described in section 2. Now, they
apply to $3+ 1$ dimensional space time. The paradox cannot be resolved
except by forbidding the initial conditions that led to them. But this
implies an influence of time machine on events, extending indefinitely
into the past. Also, such an influence implies tachyonic communications
between physical events in an era when no time machine existed
(these phenomena, which occur here in the classical regime, have been
observed by Hartle in his generalised quantum mechanics \cite{hartle}).
Moreover, the known physical laws cannot accomodate and enforce such
censorship on initial conditions which can be set at any arbitrary time
in the past in widely seperated locations.

The system under consideration is macroscopic with a classical action
$S \gg \frac{h}{2 \pi}$, the quantum action. Therefore it is unlikely that
quantising the system will resolve the paradox.
Thus, if a quantum mechanical solution does exist for
the scenario described here, it must be quite unusual, having no classical
counterpart. Also, quantisation by Hamiltonian method is not
possible in the presence of time machines since, it requires
the space time manifold to be foliable in the time direction. It is
not obvious either, how path integral method of quantisation can be carried
out since the system is macroscopic, but with {\em no} consistent
classical solution. Quantisation may be possible if, by fiat,
all the troublesome initial conditions are forbidden, but this implies
that the time machines have an influence on events, extending
indefinitely into the past, and also tachyonic communications
between physical events as described in section 2.
Thus, if the paradox in the scenario cannot be resolved, the
logical conclusion is that the time machines of the type considered here
cannot exist in $3 + 1$ dimensional space time,
maintaining the consistency of known physical laws.
This conclusion should
also be applicable to wider class of time machines since, very likely,
our scenario can be extended to them, with the consequent
irresolvability of the paradox that follows.

\vspace{2ex}

\centerline{\bf 4. Conclusion}

\vspace{2ex}

We have described a scenario first in $1 + 1$ and then in $3 + 1$
dimensional space time which leads to an irresolvable paradox in the
presence of a time machine. No consistent classical solution exists for
this scenario. Since the system is macroscopic with an action large
compared to Planck's quantum, it is unlikely that quantising the system
will resolve the paradox. Moreover, for a macroscopic system with no
consistent classical solution, it is not obvious how path integral
quantisation can be carried out
nor is a Hamiltonian method of quantisation
available in the presence of a time machine.

Forbidding by fiat all the troublesome initial conditions
will solve the problem. However, no known physical law can
enforce this kind of censorship. Also, this implies
that time machines have an influence on events, extending indefinitely
into the past, and that there will be tachyonic communications
between physical events in an era when no time machine existed.

Thus, if the above paradox cannot be resolved, the logical conclusion is
that time machines of a certain, probably large, class cannot exist in
$3 + 1$ dimensional space time, maintaining the consistency of known
physical laws.

\vspace{4ex}

Part of this work was done while S. K. R. was at
School of Mathematics, Trinity College, Dublin.
The work of S. S. is part of
a research project supported by Forbairt SC/94/218.


\vspace{2ex}

\centerline{\bf FIGURE CAPTIONS}

\vspace{2ex}

FIGURE 1: Polchinski's paradox. The figure is shown in
the X-Y plane. A ball enters the wormhole mouth W2,
emerges from W1 in the past, collides with its younger self knocking it
off its trajectory so that it does not enter W2.

FIGURE 2: Consistent solution for Polchinski's paradox.
The figure is shown in the X-Y plane. The collision
is glancing, so that the ball still
enters W2 and emerges from W1, but along a slightly different trajectory
so that the collision with its younger self is only glancing.
Dotted lines here show the original inconsistent trajectories.

FIGURE 3: Propagation of a single ball in Politzer's time machine.
The figure is shown in the X-t plane,
and the points are denoted by their $(x, t)$ coordinates. Two
typical trajectories, 1 and 2, are shown. The time machine consists of two
spatial intervals between $x = 0$ and $x = \Lambda$ at time $t = 0$
and $t = T$. The points $(y, 0_-)$ are smoothly identified with
$(y, T_+)$, and the points $(y, T_-)$ with $(y, 0_+)$, for
$0 \le y \le \Lambda$.

FIGURE 4: Propagation of two identical balls in Politzer's time machine.
The figure is shown in the X-t plane.

FIGURE 5: Paradoxical collision of a light ball L with a heavy one H
in Politzer's time machine. The figure is shown in the X-t plane,
and the points are denoted by their $(x, t)$ coordinates.
The initial velocities of L and H are $U_L$ and $U_H$ respectively.
Their initial trajectories intersect the $t = 0$ line at
$(- p', 0)$ and $(p, 0)$ respectively. The initial trajectory of H would
intersect the $t = T$ line at $(- q, T)$. L and H collide first
at A$(a, t_A) \; , \; 0 < a < \Lambda$, and H enters the time machine
at $(l, T_-)$ and emerges from it at $(l, 0_+)$ with a velocity $V_H$.
There will be a second collision between L and H at
B$(b, t_B) \; , \; b < a \; , \; t_B < t_A$, to the past of the first
collision. Here $b$ can be positive or negative, $U_H$ and $V_H$
are negative, and all other variables are positive.

FIGURE 6: The paradoxical scenario in $3 + 1$ dimensions. The figure is
shown in the X-Y plane. The points are denoted by their $(x, y)$
coordinates and the velocities by their X- and Y-components. Wormhole
mouths W1 and W2 are located respectively at $(0, h)$ and $(\Lambda, h)$.
The ball L and the disc D, of radius R,  have an initial velocity $(U, 0)$
and $(- V_x, V)$ respectively, and collide at the point A$(\Lambda, 0)$
at time $t = t_A$. After the collision, D enters W2 with a velocity
$(0, V)$ and emerges from W1 in the past with a velocity $(0, - V)$,
moving along negative Y-axis. Parametrically, it travels a distance
of $\alpha h$ with this velocity, gets slowed down to a small velocity
$(0, - \tilde{V})$ with which it travels the remaining distance
$\beta h \equiv (1 - \alpha) h$, and collides again with L
at the point B$(0, 0)$ at time $t = 0 < t_A$, to the past of the
A-collision.

\end{document}